\documentclass[twocolumn,superscriptaddress,amssymb,amsmath,nobibnotes,aps,prd,showpacs,nofootinbib]{revtex4}%
\pdfoutput=1

\usepackage{graphicx}
\usepackage{epsf}
\usepackage{bm}
\usepackage{amsmath,hyperref}
\usepackage{amsfonts}
\usepackage{amssymb}
\usepackage{epstopdf}
\usepackage{natbib}%
\usepackage{rotating}
\usepackage{pdflscape}
\usepackage{adjustbox}
\usepackage[toc,page]{appendix}
\providecommand{\U}[1]{\protect\rule{.1in}{.1in}}
\newcommand{\be}{\begin{equation}}
\newcommand{\ee}{\end{equation}}

\newcommand{\mincir}{\raise
-3.truept\hbox{\rlap{\hbox{$\sim$}}\raise4.truept\hbox{$<$}\ }}
\newcommand{\magcir}{\raise
-3.truept\hbox{\rlap{\hbox{$\sim$}}\raise4.truept\hbox{$>$}\ }}

\usepackage{color}

\begin{document}
\title{\textcolor{blue}{\small{$\quad$\hspace{1cm}PHYSICAL REVIEW D \textbf{94} (2016) 123511 }}\newline Probing the interaction between dark matter and dark energy in the presence of massive neutrinos}

\author{Suresh Kumar}
\email{suresh.kumar@pilani.bits-pilani.ac.in}
\affiliation{Department of Mathematics, BITS Pilani, Pilani Campus, Rajasthan-333031,India}

\author{Rafael C. Nunes}
\email{rcnunes@fisica.ufjf.br}
\affiliation{Departamento de F\'isica, Universidade Federal de Juiz de Fora, 36036-330,
Juiz de Fora, MG, Brazil}

\begin{abstract}

We consider the possibility of an interaction in the dark sector 
in the presence of massive neutrinos, and study the observational constraints on three different scenarios of massive neutrinos using the most recent CMB anisotropy data in combination 
with type Ia supernovae, baryon acoustic oscillations, and Hubble parameter measurements. 
When a sterile neutrino is introduced in the interacting dark sector scenario in addition to the standard model prediction of neutrinos,
we find that the coupling parameter, characterizing the interaction between dark matter and dark energy, is 
non-zero at 2$\sigma$ confidence level. The interaction model with sterile neutrino is also found 
to be a promising one to alleviate the current tension on Hubble constant. We do not find the evidence for a coupling in the dark sector when the possibility of a sterile neutrino is discarded.

\end{abstract}

\maketitle

\section{Introduction}
\label{sec:intro}
Since the discovery of accelerated expansion of Universe, there have been numerous efforts on theoretical and
observational grounds to explain cause of the cosmic acceleration. The most popular theoretical model, 
which explains the cosmic acceleration and fits well to the currently available observational data, is the standard
$\Lambda$CDM (cosmological constant $\Lambda$ + cold dark matter) model \cite{Planck2015}. 
In this model, $\Lambda$ is a candidate of dark energy (DE) which is believed 
to be responsible for the current accelerated expansion of the Universe. This model, 
however, suffers from some theoretical problems \cite{Peebles03, Copeland06}, which  motivated the researchers to 
propose new models of DE. One such problem is the coincidence problem which refers to the fact that 
there is no explanation for the same order of the energy densities of dark matter (DM) and vacuum energy today. 
To alleviate the coincidence problem, the interaction between DM and DE components 
of the Universe has been proposed/studied in the literature (see \cite{DE_DM_1,DE_DM_2} for recent reviews). 
The interaction in the dark sector is quite appealing since DM and DE are the dominant components in the 
overall energy budget of the Universe. Moreover, it has recently been shown that the current observational 
data can favor the late-time interaction
between DM and DE \cite{Salvatelli:2014zta,Sola:2015,Richarte:2015maa,Valiviita:2015dfa,Murgia:2016ccp}.

Following \cite{coupled01,coupled02,coupled03}, an interacting dark sector scenario has been investigated with the latest observational data in a recent paper \cite{coupled04}, where DM and DE are 
allowed to interact via a free coupling parameter (more details are given in the next section). In the present study, we shall consider this cosmological scenario of coupling in the dark sector along with massive neutrinos as the physics of neutrinos has important implications on the formation of the large scale structure, 
big bang nucleosynthesis, cosmic microwave background (CMB), and in other cosmological 
information sources (see \cite{Dolgov,Lesgourgues} for review). Also, see \cite{abaz15} for a recent report on neutrinos studies. The sum of the active neutrino 
masses ($\sum m_{\nu}$) as well as the effective number of relativistic degrees of freedom ($N_{\rm eff}$), 
are still the unknown quantities, and these have received the attention of researchers in different cosmological contexts lately. 
For instance, within the framework of $\Lambda$CDM model, Planck collaboration \cite{Planck2015}  has measured $\sum m_{\nu} < 0.49$ eV (from Planck
TT, TE, EE + lowP) and $N_{\rm eff} = 3.04 \pm 0.33$ at 2$\sigma$ confidence level (CL). Constraints on neutrino masses are also 
investigated from baryonic acoustic oscillations (BAO) \cite{nu1}, and  the
spatial distribution of DM haloes and galaxies \cite{nu2}. Evidence for massive neutrinos is found via 
CMB and lensing measurements in \cite{nu3}. Forecasting of the measurements of neutrino masses with future cosmological data 
is addressed in \cite{nu4}. Massive neutrinos have also been studied in $f(R)$ gravity \cite{fR_nu1,fR_nu2}, 
holographic DE \cite{hde_nu}, parametric dynamical DE \cite{cc} and scalar field DE  \cite{sf_nu,sf_nu2,sf_nu3} models. 

Given the motivation and the recent indication in favor of an interaction in the dark sector, the main target of this paper is to explore
the interacting dark sector scenario, investigated recently in \cite{coupled04}, with massive neutrinos (including the possibility of a sterile neutrino)
and its possible degeneracy with the coupling parameter by using the latest observational data. 
Massive neutrinos may generate significant effects on some cosmological parameters. It would be useful to 
explore new perspectives in the direction of a possible interaction in the dark sector. The paper is organized as follows. In the next section, we briefly review the coupled DM and DE 
scenario in the presence of the massive neutrinos. In section III, we present the data sets 
used and the statistical results. We summarize findings of the study in the final section. 

\section{Interaction in the dark sector with massive neutrinos}
\label{sec:model}

We consider a spatially-flat Friedmann-Robertson-Walker (FRW) Universe, and 
assume that DM and DE are not conserved separately. 
The energy conservation law for the interacting DM-DE components 
reads as 
\[\nabla_{\mu}T^{\mu \nu}_{\rm dm}=-\nabla_{\mu}T^{\mu \nu}_{\rm de}.\]
It leads to
\begin{align}
\label{Q_equation}
\dot{\rho}_{\rm dm} + 3\frac{\dot{a}}{a}\rho_{\rm dm} = -\dot{\rho}_{\rm de} -
3\frac{\dot{a}}{a}\left(\rho_{\rm de} + p_{\rm de}\right) = Q,
\end{align}
where $Q$ is the interaction function between DM and DE, being that for $Q > 0$, 
the energy flow takes place from DE to DM, and for $Q < 0 $, the energy flows from DM to DE.
\\

The Friedmann equation can be written as
\begin{align}
\label{friedmann}
3 H^2 = 8 \pi G \left(\rho_{\gamma} + \rho_{\nu} + \rho_{\rm b} + \rho_{\rm dm} +
\rho_{\rm de}  \right),
\end{align}
where $H= \dot{a}/a$ is the Hubble parameter with an over dot denoting
derivative with respect to the cosmic time. Further $\rho_{\gamma}$, $\rho_{\nu}$, $\rho_{\rm b}$,
$\rho_{\rm dm}$, and $\rho_{\rm de}$ stand for the energy densities of photons, neutrinos, baryons, cold DM and DE, respectively. As usual, a subindex zero attached to any quantity shall mean that it is evaluated at the present time.
\\

Let us consider that the DM particles can undergo dilution throughout the cosmic history with a small deviation from the standard evolution 
characterized by a constant $\delta$ as follows

\begin{align}
\rho_{\rm dm} = \rho_{\rm dm,0}\,a^{-3 + \delta},
\label{rho0a3}
\end{align}
and consider that the DE is described by an equation of state parameter $w = - p_{\rm de}/\rho_{\rm de}=const.$, being $w > -1$ featuring 
a quintessence field, and $w < -1$ a phantom field. From the eq. (\ref{Q_equation}) and (\ref{rho0a3}) we find
\begin{align}
\rho_{\rm de} = \rho_{\rm de,0}\,a^{-3(1+w)} + \frac{\delta\, \rho_{\rm dm,0}}{3|w| - \delta}\left[ 
a^{-3 +\delta}- a^{-3(1+w)} \right].
\end{align}

The parameter $\delta$ is the coupling constant that characterizes the interaction between DM and DE via the interaction term $ Q = \delta H \rho_{\rm dm} $, which can be obtained by  using (\ref{rho0a3}) into  (\ref{Q_equation}). Obviously, $\delta=0$ implies 
the absence of interaction while $\delta<0$ corresponds to the energy transfer from DM to DE. 
Hence, the
parameter $\delta$ quantifies the deviation from the non-interacting
case.

\begin{figure}\centering
\includegraphics[width=9.0cm]{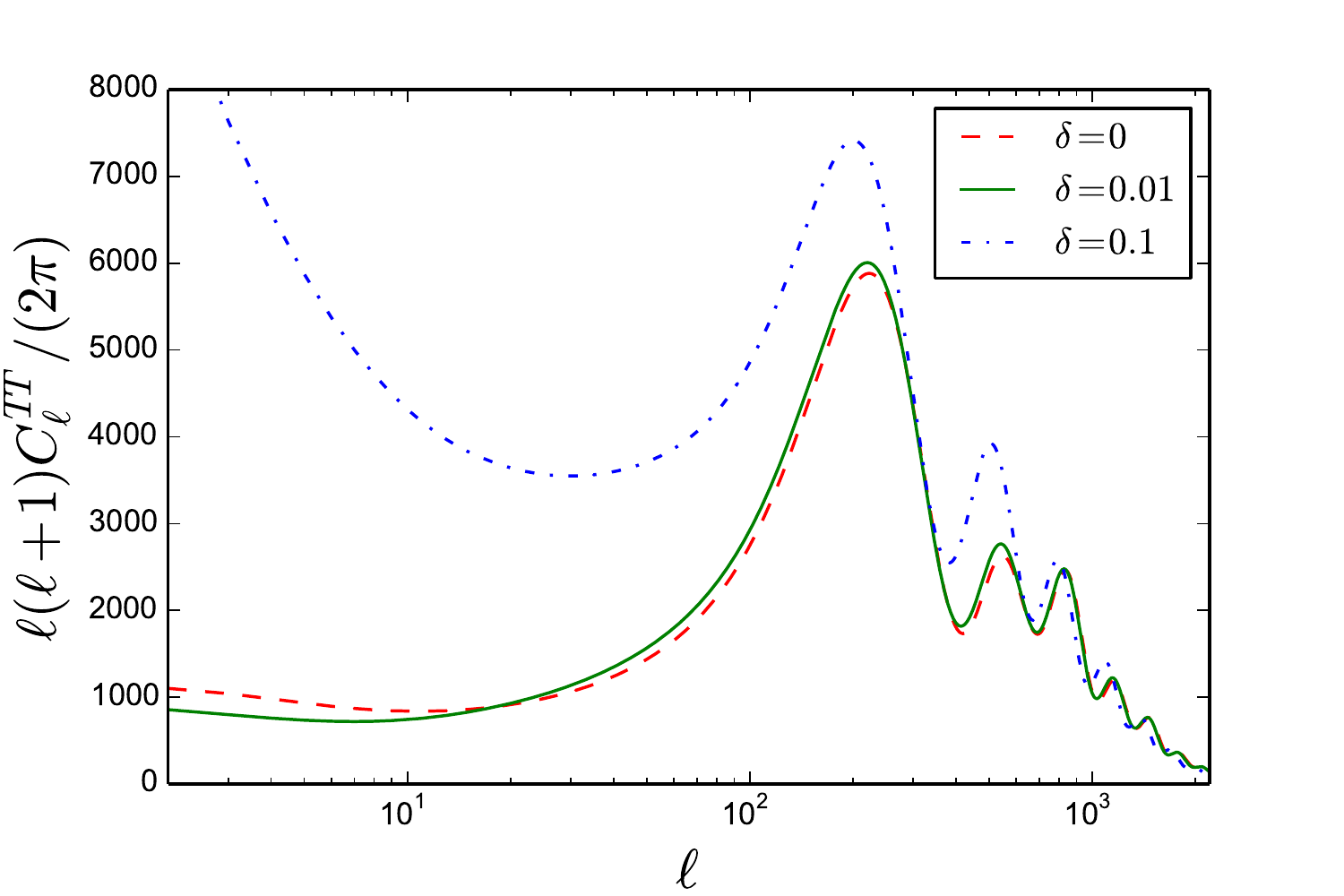}
\caption{\label{cmb_tt}{\it{The figure shows the theoretical predictions for angular power spectrum of the CMB temperature anisotropy 
for  some specific values of the dark sector coupling parameter $\delta$, where the other parameters are fixed to their mean values given in table \ref{tab1}. }}}
\end{figure}

The above scenario 
was originally introduced in \cite{coupled01}. Later, it was generalized and analyzed with observational data 
in \cite{coupled02,coupled03,coupled04}.
It has been argued in \cite{pert01} that simple interaction between DM and DE with constant $w$ can generate 
instability in the dark sector perturbations at early times when the coupling function is proportional to 
DM energy density: the curvature perturbation 
blows up on super Hubble scales. However, the perturbations could be stable if the coupling is proportional to the energy density of DE \cite{pert02}. Within the synchronous gauge (used in this work), it is possible to calculate the effects on CMB 
at all angular scales for $\delta \ll 1$. For instance, see Figure \ref{cmb_tt}, where the case $\delta \lesssim 10^{-2}$ 
is observed with only small deviation in comparison to the non-coupled case, i.e., $\delta = 0$). 
That is a good approximation, since we expect the interaction between the dark components to be weak, 
as confirmed in some other investigations \cite{Valiviita:2015dfa,coupled06,coupled07,coupled08}.
At the level of perturbations,  there are the difficulties in properly defining the dark sector interaction from a phenomenological perspective as demonstrated in \cite{coupled09}.

Let us now review how coupling in the dark sector affects the linear perturbation evolution. 
The most general scalar mode perturbation is defined by the following metric \cite{Mukhanov, Malik}

\begin{eqnarray}
ds^2 = -(1+ 2 \phi)^2 dt^2 + 2 a \partial_i B dt dx + \nonumber \\  
a^2 [(1-2 \psi) \delta_{ij} + 2 \partial_i \partial_j E] dx^i dx^j.
\end{eqnarray}

Here, we follow \cite{delta_cdm,delta_cdm2}, where a synchronous gauge has been 
adopted, i.e.,  $\phi = B = 0$, $\psi = \eta$, and $k^2 E = - h/2 - 3 \eta$. The energy and momentum conservation 
equations for each fluid component in synchronous gauge are given by

\begin{eqnarray}
\label{delta_dm1}
\dot{\delta}_{i} + 3H(\delta \rho_i + \delta p_i ) - 3(\rho_i + p_i)\dot{\psi} + \nonumber \\ 
(\rho_i + p_i) \frac{k^2}{a^2}(\theta_i + \sigma) = \delta Q_i + Q_{i}
\end{eqnarray}
and 

\begin{eqnarray}
\label{theta_1}
\dot{\theta_i}(\rho_i + p_i) -3 c_s^2 H (\rho_i + p_i) \theta_i + (\rho_i + p_i) \phi + \nonumber \\  
+ \delta p_i + \frac{2}{3}\frac{k^2}{a^2}\Pi_i = f_i + Q_i \theta_i - (1+c_s^2)Q_i \theta_i,
\end{eqnarray}
where in the above equantions, the quantities $\sigma$, $\Pi$, $\delta \rho$, $\delta p$, $\theta$, $c^2_s$, are the shear, 
total anisotropic stress, density perturbation, pressure pertubation, velocity pertubation, and the adiabatic sound speed, respectively. 
The functions $\delta Q_i$ and $f_i$ refer to the energy transfer and momentum transfer, respectively. 
\\

We assume that the DE perturbation in the DM comoving frame is 
identically zero, i.e., $\delta_{\rm de} =0$. Hence, in an interacting DE-DM scenario, from eqs. 
(\ref{delta_dm1}) and (\ref{theta_1}), the perturbed parts of the energy and momentum for the DM evolution respectively read as
\cite{delta_cdm,delta_cdm2}:

\begin{eqnarray}
\label{delta_dm2}
\dot{\delta}_{\rm dm} - \frac{Q}{\rho_{\rm m}}\delta_{\rm dm} - \frac{k^2}{a^2}\theta_{\rm dm} + 
\frac{\dot{h}}{2} = 0,
\end{eqnarray}

\begin{eqnarray}
\label{theta_2}
\dot{\theta}_{\rm dm} = \frac{Q}{\rho_{\rm dm}} \theta_{\rm dm}.
\end{eqnarray}

In the above equations, we have neglected the shear stress and the adiabatic sound speed of the DM, which is always 
negligible because of its non-relativistic character. In synchronous gauge the DM velocity is zero. 
The baryons, photons and massive neutrinos are conserved independently, and the perturbation equations 
follow the standard evolution described in \cite{Ma_Bertschinger}.
\\

In the next section, we discuss the observational constraints on the free parameters of 
the model under consideration in the presence of massive neutrinos. On the basis of neutrino 
masses, we consider the following models.
\\

\noindent\textbf{Model I}. We consider a model with three active neutrinos \footnote{Three active neutrinos
represent a base model in the literature that features two massless and one massive neutrino.} 
subject to the condition that $N_{\rm eff}=3.046$ \cite{Planck2015}. It is the case, normally considered in the literature to investigate the effects 
of $\sum m_{\nu}$ on a particular cosmological scenario. The base parameters set for the Model I is 

\begin{eqnarray*}
\label{P1}
P = \{100\omega_{\rm b}, \, \omega_{\rm dm},  \, 100\theta_{s}, \, \ln10^{10}A_{s}, \, \nonumber \\ 
n_s, \, \tau_{\rm reio},  \, \sum m_{\nu}, \, \delta, \, w \}
\end{eqnarray*}

\noindent\textbf{Model II}. We take the Model I with the effective number of relativistic species as a free parameter, i.e, 
Model I + $N_{\rm eff}$. The base parameters set for the Model II is

\begin{eqnarray*}
\label{P1}
P = \{100\omega_{\rm b}, \, \omega_{\rm dm}, \, 100\theta_{s}, \, \ln10^{10}A_{s}, \, \nonumber \\ 
n_s, \, \tau_{\rm reio}, \, N_{\rm eff}, \, \sum m_{\nu}, \, \delta, \, w \}
\end{eqnarray*}
\\

\noindent\textbf{Model III}. The Planck collaboration fixes the mass ordering of the active neutrinos 
to the normal hierarchy with the minimum masses allowed by oscillation experiments, $m_1 = m_2 = 0$ eV and $m_3 = 0.06$ eV. 
Any excess mass is considered to be from a single additional mass state $m_4$, which can be related 
to the sterile neutrino mass, $m_{{\nu}_s}$. Taking this into account, 
the mass splitting with relation to the sterile neutrino can be written as $\Delta m_{41} = m_4$.
In order to constrain the sterile neutrino mass, the effective number of neutrino species is $N_{\rm eff} = 4.046$, 
where we have a contribution of 3.046 via standard model prediction and plus 1 from 
the degree of freedom associated with the sterile neutrino. These considerations are 
used in our analysis, i.e., we include in the base model, two massless neutrinos, one massive
neutrino with minimum mass (0.06 eV) and one sterile neutrino with full
thermalisation and mass $m_{{\nu}_s}$. The base parameters set for the Model III is
 
\begin{eqnarray*}
 \label{P2}
 P = \{100\omega_{\rm b}, \omega_{\rm dm}, 100\theta_{s}, \ln10^{10}A_{s}, n_s, \nonumber \\ 
 \tau_{\rm reio}, \,  m_{{\nu}_s},  \, \delta, \, w \}.
\end{eqnarray*}

One may notice that the above three models carry free parameters in addition to the 
six parameters $100\omega_{\rm b}$, $\omega_{\rm dm}$, $100\theta_{s}$, $\ln10^{10}A_{s}$, $n_s$ and $\tau_{\rm reio}$ of the $\Lambda$CDM cosmology. 
Therefore, these models are simple extensions of the standard $\Lambda$CDM model.

\section{Data sets and Results}

In what follows, first we briefly describe the observational data sets used to constrain the parameters of the models under consideration.
\\

\noindent\textbf{Planck CMB}: We use the full Planck 2015 data \cite{Planck2015} comprised of temperature (TT), polarization (EE) and the cross correlation 
of temperature and polarization (TE) together with the CMB lensing power spectrum. 
\\

\noindent\textbf{JLA}: We use the latest ``joint light curves" (JLA) sample \cite{snia3}, comprised of 740 type Ia supernovae
in the redshift range $z \in [0.01, 1.30]$. 
\\

\noindent\textbf{BAO}: We use the BAO measurements from the  Six  Degree  Field  Galaxy  Survey  (6dF) \cite{bao1}, 
the  Main  Galaxy  Sample  of  Data  Release 7  of  Sloan  Digital  Sky  Survey  (SDSS-MGS) \cite{bao2}, 
the  LOWZ  and  CMASS  galaxy  samples  of  the Baryon  Oscillation  Spectroscopic  Survey  (BOSS-LOWZ  and  BOSS-CMASS,  
respectively) \cite{bao3},  and the distribution of the LymanForest in BOSS (BOSS-Ly) \cite{bao4}.  
\\

\noindent\textbf{CC+$H_0$}: We use the cosmic chronometers (CC) data set comprising of 30 measurements 
spanning the redshift range $0 < z < 2$,  recently compiled in \cite{cc} . Furthermore, we also include the recently measured new local
value of Hubble constant given by $H_0=73.24 \pm 1.74$   km
s${}^{-1}$ Mpc${}^{-1}$ as  reported in \cite{riess}. 
\\

We modified the publicly available CLASS \cite{class} and Monte Python \cite{monte} codes for the models 
under consideration, and constrained the free parameters of the models by utilizing two different combinations of data sets: Planck CMB + JLA + BAO and
Planck CMB + JLA + BAO + CC + $H_0$. We used Metropolis Hastings algorithm with uniform priors on the model parameters 
to obtain correlated Markov Chain Monte Carlo (MCMC) samples from CLASS/Monte Python code, and analyzed these 
samples by using the GetDist Python package \cite{antonygetdist}.
\\

\begin{table*}[ht]
\caption{\label{tab1} Constraints on the free parameters and some derived parameters $H_0$, $\sigma_8$ (RMS matter fluctuations today in linear theory) and $\Omega_{\rm m}$ (Matter density including massive neutrinos today divided by the critical density) of the three models considered in the present study. Mean values of the parameters are displayed with 1$\sigma$ and 2$\sigma$ errors.  First row entries against each parameter show the constraints from Planck CMB + JLA + BAO data, and the second row entries for each parameter display the constraints 
from Planck CMB + JLA + BAO + CC + $H_0$ data. The parameter $H_0$ is in the units of km
s${}^{-1}$ Mpc${}^{-1}$, while $\sum m_{\nu}$ and $m_{{\nu}_s}$ are in the units of eV.}
     \begin{center}
\begin{tabular} { l  l l l}
\hline
 Parameter          & Model I & Model II & Model III \\
 \hline
 \hline
{$100\omega_{\rm b }$} & $2.223^{+0.017 +0.033}_{-0.017 -0.033}$  & $2.220^{+0.025 +0.059}_{-0.032 -0.056}$ & $2.329^{+0.019+0.041}_{-0.019-0.037}$   \\
                   & $2.220^{+0.015 -0.032}_{-0.015 -0.029}$  & $2.235^{+0.023 +0.052}_{-0.028 -0.048}$ & $2.327^{+0.017+0.033}_{-0.017-0.032}$  \\
                                     
{$\omega_{\rm dm }  $} & $0.1178^{+0.0017 +0.0033 }_{-0.0017 -0.0034}$ & $0.1172^{+0.0030 +0.0075}_{-0.0039 -0.0072}$ & $0.1353^{+0.0021+0.0040}_{-0.0021-0.0039}$ \\
                   & $0.1184^{+0.0018 +0.0038}_{-0.0020 -0.0036}$  & $0.1204^{+0.0032 +0.0064}_{-0.0032 -0.0060}$ & $0.1356^{+0.0020+0.0039}_{-0.0020-0.0038}$ \\
                                   
{$100\theta_{s }  $} & $1.04195^{+0.00030 +0.00060}_{-0.00030 -0.00059}$ & $1.0421^{+0.0006 +0.0011}_{-0.0005 -0.0011}$ & $1.0399^{+0.0003+0.0006}_{-0.0003-0.0006}$\\
                     & $1.04196^{+0.00031 +0.00069}_{-0.00039 -0.00063}$ & $1.0417^{+0.0004 +0.0010}_{-0.0005 -0.0009}$ & $1.0399^{+0.0002+0.0006}_{-0.0003-0.0005}$\\

{$\ln10^{10}A_{s }$} & $3.082^{+0.028 +0.055}_{-0.028 -0.051}$ & $3.078^{+0.029 +0.058}_{-0.033 -0.056}$ & $3.142^{+0.030+0.060}_{-0.030-0.056}$\\
                     & $3.077^{+0.028 +0.056}_{-0.028 -0.054}$ & $3.080^{+0.028 +0.061}_{-0.033 -0.058}$ & $3.142^{+0.030+0.060}_{-0.030-0.056}$\\
                   
{$n_{s }         $} & $0.9680^{+0.0047 +0.0091}_{-0.0047 -0.0092}$ & $0.9663^{+0.0073 +0.0180}_{-0.0096 -0.0160}$ & $1.0003^{+0.0053+0.0110}_{-0.0053-0.0100}   $\\
                    & $0.9662^{+0.0047 +0.0084}_{-0.0040 -0.0093}$ & $0.9706^{+0.0076 +0.0150}_{-0.0076 -0.0150}$ & $0.9994^{+0.0046+0.0094}_{-0.0046-0.0090}$ \\
                                 
{$\tau_{\rm reio }   $} & $0.075^{+0.015 +0.029}_{-0.015 -0.029}$ & $0.075^{+0.015 +0.028}_{-0.015 -0.028} $ &$0.090^{+0.015+0.039}_{-0.020-0.032} $\\
                    & $0.073^{+0.015 +0.030}_{-0.015 -0.028}$ &$0.072^{+0.014 +0.031}_{-0.017 -0.028} $  & $0.089^{+0.016+0.031}_{-0.016-0.030} $\\
                        
{$N_{\rm eff}  $}       & $ 3.046 $   & $1.99^{+0.16 +0.41}_{-0.22 -0.38}$  & $ 4.046 $\\
                    & $ 3.046 $   & $2.15^{+0.17 +0.35}_{-0.17 -0.35}$ & $ 4.046 $ \\
                                                                                             
{$\sum m_{\nu}  $} & $ < 0.15 \, (<0.28)$  & $<0.10 \, (<0.24) $  & $0.06$ \\
                   & $<0.18   \, (<0.32) $ & $<0.17 \,  (<0.28) $ & $0.06$ \\

{$ m_{{\nu}_s}  $}  &$ --$     &   $--$ &    $ < 0.19 \, (< 0.40)$                                    \\
                    & $--$    &   $--$ &     $<0.25 (< 0.48) $                                  \\

{$\delta         $} &$-0.0008^{+0.0011 +0.0021}_{-0.0011 -0.0020}$  &$-0.0008^{+0.0012 +0.0025}_{-0.0013 -0.0025}$ & $0.0022^{+0.0011+0.0021}_{-0.0011-0.0021}$\\
                    & $-0.0006^{+0.0011 +0.0023}_{-0.0012 -0.0023}$ & $-0.0002^{+0.0012 +0.0023}_{-0.0012 -0.0024}$ &$0.0023^{+0.0011+0.0022}_{-0.0011-0.0021}$ \\
                    
{$w        $} & $-0.973^{+0.039 +0.073}_{-0.039 -0.079}$ & $-0.972^{+0.038 +0.076}_{-0.038 -0.073}$  & $-0.956^{+0.040+0.078}_{-0.040-0.072}  $\\
              & $-0.932^{+0.035 +0.072}_{-0.035 -0.067}$ & $-0.938^{+0.036 +0.074}_{-0.036 -0.069}$  & $-0.950^{+0.036+0.073}_{-0.036-0.071}  $\\
          
\hline
{$H_0            $} &$68.4^{+1.2 +2.3}_{-1.2 -2.2}$ &$68.4^{+1.5 +3.0}_{-1.5 -3.0} $  & $73.4^{+1.2+2.3}_{-1.2-2.3}        $\\
                    & $69.9^{+0.9 +1.9}_{-1.1 -1.8}$ & $70.3^{+1.1 +2.0}_{-1.1 -2.0}$ & $73.2^{+1.0+1.9}_{-1.0-1.8}               $\\

{$\sigma_8        $} & $0.816^{+0.016 +0.026}_{-0.013 -0.029}$ & $0.818^{+0.016 +0.028}_{-0.013 -0.029} $ & $0.835^{+0.016+0.031}_{-0.014-0.031}$ \\
                     & $0.822^{+0.014 +0.023}_{-0.011 -0.026}$  & $0.825^{+0.014 +0.027}_{-0.014 -0.028} $ & $0.829^{+0.022+0.034}_{-0.015-0.039}   $\\
                                         
{$\Omega_{\rm m }    $} & $0.302^{+0.010 +0.020}_{-0.010 -0.019}$ & $0.300^{+0.009 +0.020}_{-0.009 -0.018} $ & $0.299^{-0.010+0.017}_{-0.010-0.018}   $ \\
                    & $0.291^{+0.008 +0.015}_{-0.008 -0.016}$ & $0.292^{+0.008 +0.018}_{-0.009 -0.016} $ &  $0.302^{+0.008+0.015}_{-0.008-0.015}   $\\
                 
\hline

{$\chi^2_{min}/2$}         &  $6820.38$     &   $6820.83$         &  $6832.08$  \\
                           &  $6832.16$     &   $6830.49$         &  $6842.10$  \\
\hline
\end{tabular}
\end{center}

\label{tab1}
\end{table*}
\begin{figure}\centering
\includegraphics[width=9.0cm]{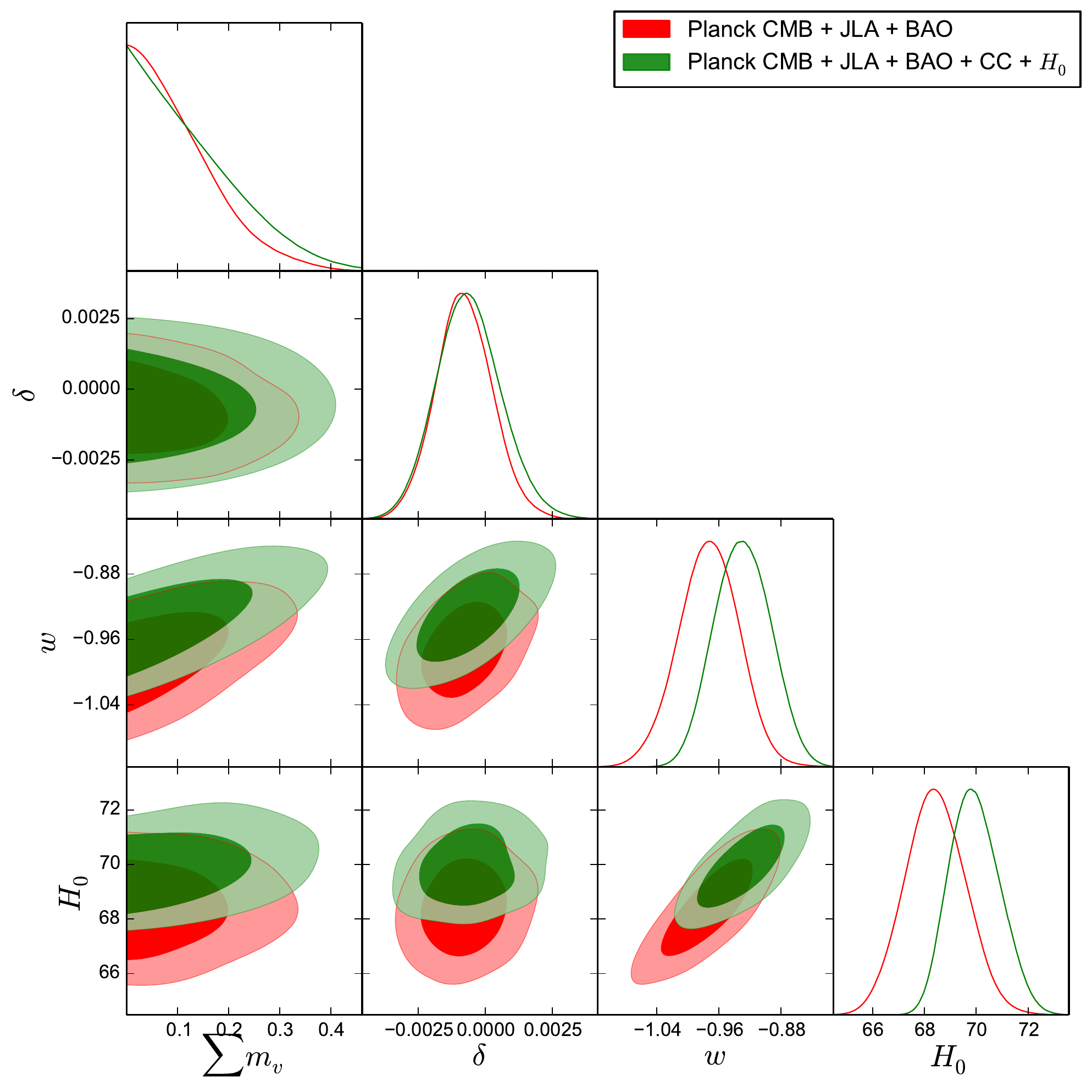}
\caption{\label{contour1} {\it{One-dimensional marginalized distribution, and  1$\sigma$ and 2$\sigma$ two-dimensional confidence contours for some selected parameters of the Model I. }}}
\end{figure}

\begin{figure}
\includegraphics[width=9.0cm]{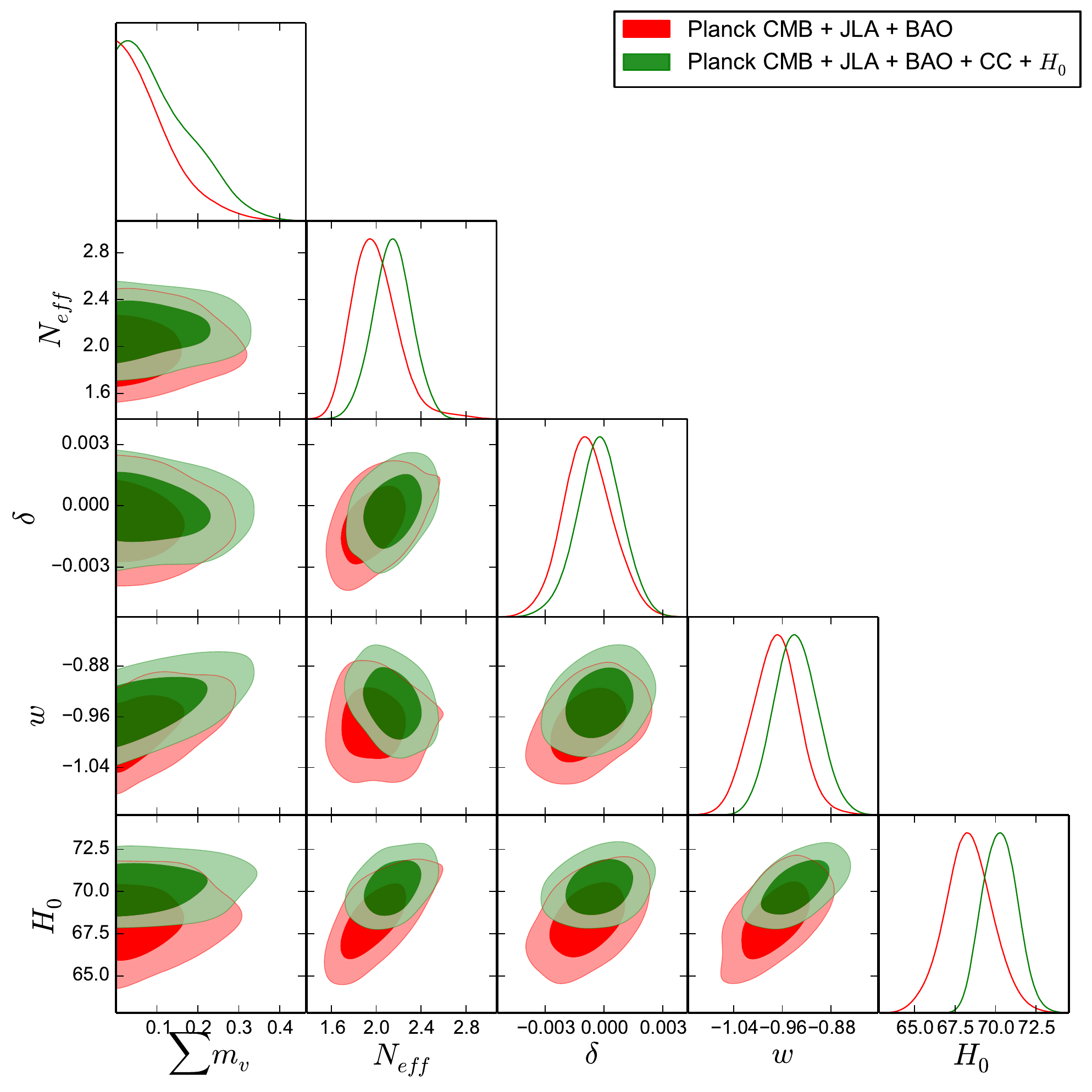}
\caption{\label{contour2} {\it{One-dimensional marginalized distribution, and  1$\sigma$ and 2$\sigma$ two-dimensional confidence contours for some selected parameters of the Model II.}}}
\end{figure}

\begin{figure}
\includegraphics[width=9.0cm]{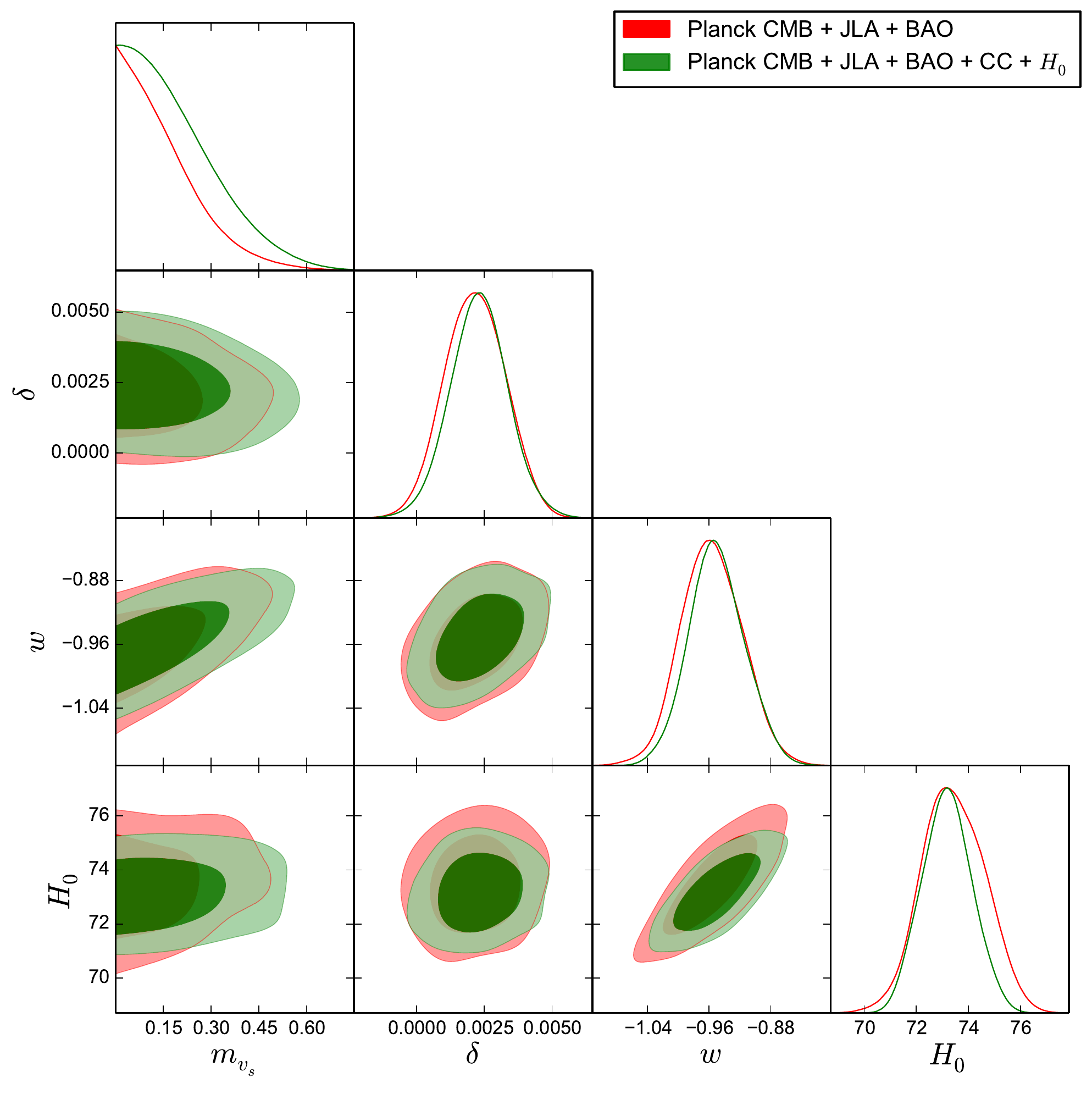}
\caption{\label{contour3} {\it{ One-dimensional marginalized distribution, and  1$\sigma$ and 2$\sigma$ two-dimensional confidence contours for some selected parameters of the Model III.}}}
\end{figure}

Table \ref{tab1} summarizes the main results of the statistical analysis
carried out using two different combinations of data sets: Planck CMB + JLA + BAO and Planck CMB + JLA + BAO + CC + $H_0$.
\\

First, we discuss the results for the Model I. Figure \ref{contour1} 
shows the contour plots for some selected parameters of the Model I
considering two combined data sets: Planck CMB + JLA + BAO and Planck CMB + JLA + BAO + CC + $H_0$, showing the change in constraints, if any, on the model parameters in the absence or presence of the data set CC + $H_0$.
We note that $\sum m_{\nu} < 0.28 \, (32) $ eV at 
2$\sigma$ confidence level (CL) in the presence of the interaction in the dark sector characterized 
by a coupling constant  $\delta \simeq 0$ in both cases. 
On the other hand, the equation of state of the DE is $w > -1$ at $\sim$ 1.8$\sigma$ CL in the joint analysis, indicating a quintessence character of the DE field. Recently similar results for $\delta$ have 
been reported in \cite{coupled04}, but with DE favored by a phantom field. 
Here it deserves to mention that the presence of massive neutrinos was not considered in \cite{coupled04}, 
and the full Planck data set as well. Therefore, the present investigation makes the present status of 
the model more accurate and realistic with the current cosmological data.  
\\

Figure \ref{contour2} shows the contour plots for some selected parameters of the Model II.
We do not find evidence of interaction between DM and DE in this model as well since the coupling parameter $\delta \simeq 0$ at 1$\sigma$ CL. We note that $\sum m_{\nu} < 0.24 \, (0.28)$ eV
and $N_{\rm eff}=1.99^{+0.41}_{-0.38} \, (2.15^{+0.35}_{-0.35})$ for Planck CMB + JLA + BAO (Planck CMB + JLA + BAO + CC + $H_0$) data, 
respectively. Our results on $N_{\rm eff}$ characterize the contribution only of relativistic relics since we have considered one free massive neutrino in our analysis, which is standard in the literature. Thus, 
the total contribution of the species is the best fit for $N_{\rm eff}$ plus 1, 
which, as expected,  is around 3 (3.15) for Planck CMB + JLA + BAO (Planck CMB + JLA + BAO + CC + $H_0$) data. Taking into account the joint analysis result and the active massless neutrinos, we note 
$\Delta N_{\rm eff} = N^{\rm best \, fit}_{\rm eff} -2\sim 0.15$. 
We note that in the joint analysis, we have $w > -1$ at $\sim 1.8 \sigma$ CL.
We observe positive correlation between $H_0$ and $N_{\rm eff}$ in Figure \ref{contour2}, and consequently it is easy to see that $H_0 > 70.0$ km  for $\Delta N_{\rm eff} > 0.15$.
s${}^{-1}$ Mpc${}^{-1}$. Thus, it seems to reconcile the tension between the local and global measurements of the Hubble constant\footnote{Assuming standard $\Lambda$CDM the Planck
data gives $H_0 = 67.27 \pm 0.66$     km s${}^{-1}$ Mpc${}^{-1}$ that is about
two standard deviations away from the value $H_0=73.24 \pm 1.74$   km s${}^{-1}$ Mpc${}^{-1}$ as  reported in \cite{riess}. See \cite{Valentino16,Huang16} for some recent studies in this regard.}. 
\\

In Model III, the inclusion of one sterile neutrino leads to $N_{\rm eff} = 4.046$, i.e., 
a variation of $\Delta N_{\rm eff} = 1$ in the standard model prediction. Figure \ref{contour3} 
shows the contour plots for some selected parameters of the Model III. We note that $m_{{\nu}_s}$ $ \, < 0.40 \, (< 0.48)$ at 2$\sigma$ CL 
for Planck CMB + JLA + BAO (Planck CMB + JLA + BAO + CC + $H_0$) data. On the other hand, contrary to what we observed in the analysis of the Models I and II, here in the joint analysis we note that the coupling parameter $\delta $ appears with a significant positive value. More specifically, we have $0.0002 \leq \delta \leq 0.0045$ at 2$\sigma$ CL, evidencing a possible interaction 
in the dark sector within of the framework of Model III.
Also, we note that equation of state parameter $w$ of the DE is greater than $ -1$ at a CL little more than 1$\sigma$.
Another important point to be noted here is that the inclusion of one sterile neutrino besides the standard model prediction of neutrinos,
yields $H_0 = 73.2$ km s${}^{-1}$ Mpc${}^{-1}$ (the result from the joint analysis), thus leading to reconcile the current tension on $H_0$. 
Therefore, in addition to providing an evidence for the interaction between DM and DE at 2$\sigma$ CL, 
the Model III can also be useful to alleviate the $H_0$ tension.
\\

It has recently been shown that an interaction in the dark sector can reconcile the $\sigma_8$ tension between 
CMB and structure formation measurements \cite{Pourtsidou}. Therefore, it is expected that interaction in the dark sector in 
the presence of massive neutrinos can reconcile both the tensions ($H_0$ and $\sigma_8$) using more sophisticated phenomenological models 
(including another interaction term $Q$ as well as non-cold dark matter relics) in combination
with CMB and structure formation data, besides the standard cosmological tests. 
A comprehensive analysis including these considerations shall be presented in a forthcoming paper \cite{SK}.

\section{Final Remarks}

In this work, we have investigated a cosmological scenario where DM and DE are allowed to have mutual interaction via the free 
coupling parameter $\delta$ while assuming constant equation of state parameter $w$ for the DE. We have constrained 
this interaction in the dark sector in the presence of massive neutrinos
using the latest observational data from Planck CMB, JLA, BAO, CC, and $H_0$ by considering three models. 
In first model (Model I), three active neutrinos are considered with $N_{\rm eff}=3.046$ while the second model (Model II) is being the Model I + $N_{\rm eff}$. The third model (Model III) 
includes the presence of one sterile neutrino. The results of the statistical analysis are displayed in 
Table \ref{tab1} for the three models. The main findings of the study are summarized in the following.

\begin{itemize}

\item In Models I and II, we observe that the mean value of the coupling parameter is very close to 0, i.e., $\delta \simeq 0$. 
This shows that interaction in the dark sector is not favored by the present observational data within the 
framework of these two models. In Model I, we obtain $\sum m_{\nu} < 0.32$ eV, while the masses of the active neutrinos in Model II are constrained to $\sum m_{\nu} < 0.28$ eV with $\Delta N_{\rm eff} \sim 0.15$ at 2$\sigma$ CL.

\item The equation of state parameter of the DE in all the three models is observed to have values greater than $-1$ at around 1.8$\sigma$ 
CL  in the joint analysis, indicating the quintessence character of DE field in the models under consideration.

\item In Model III, we have found that the coupling parameter is constrained as $0.0002 \leq \delta \leq 0.0045$ 
at 2$\sigma$ CL from the joint analysis, featuring a favor of the considered data  for an interaction in the dark sector. The sterile neutrino is 
observed to have $m_{{\nu}_s} < 0.48$ eV via Planck CMB + JLA + BAO + CC + $H_0$. 
We note that $H_0 = 73. 4$ km s${}^{-1}$ Mpc${}^{-1}$ ($73.2$ km s${}^{-1}$ Mpc${}^{-1}$) from Planck CMB + JLA + BAO (Planck CMB + JLA + BAO + CC + $H_0$) data. Therefore, the Model III is a promising model to alleviate
the current tension between local and global determinations of the Hubble constant. 

\end{itemize}

A mutual interaction between DM and DE has been investigated in different contexts, 
and it has recently been shown to have a strong potential to be an extension of the standard $\Lambda$CDM cosmology
\cite{Salvatelli:2014zta,Sola:2015,Richarte:2015maa,Valiviita:2015dfa,Murgia:2016ccp}. 
Here we have shown how the inclusion of massive neutrinos
can constrain a cosmological scenario of coupled dark sector represented by the interaction 
function $Q= \delta H \rho_{\rm dm}$ with DE described by $w = const.$, using the full CMB anisotropy data, type Ia supernovae, 
BAO and Hubble parameter measurements. We have shown that the presence of sterile neutrino
can provide new perspectives in this context such as  a possible indication of interaction in the dark sector. 
Obviously, other phenomenological models of coupled DE are also worth exploring with massive neutrinos.
\section*{Acknowledgments}
S.K. gratefully acknowledges the support from Science
and Engineering Research Board, Department of Science \&
Technology  (SERB-DST), Government of India Project
No. EMR/2016/000258.

\end{document}